\documentclass[conference]{IEEEtran}
    \IEEEoverridecommandlockouts
    \usepackage{amsmath}
\usepackage{amssymb}
\usepackage{amsthm}
\usepackage{hyperref}

\newcommand{\hs}{\mathcal{H}}

\newcommand{\cmss}{\fontfamily{cmss}\selectfont}

\newcommand{\A}{\mathcal{A}}
\newcommand{\B}{\mathcal{B}}
\newcommand{\CH}{\mathcal{CH}}

\newtheorem{theorem}{Theorem}
\newtheorem{lemma}{Lemma}

\usepackage{array}
\usepackage{setspace}
\usepackage{graphicx}
\usepackage{nccmath}
\usepackage[ruled,noline,linesnumbered]{algorithm2e}

\begin{document}

\title{Adamastor: a New Low Latency and Scalable Decentralized Anonymous Payment System}

\author{Rui Morais,
        Paul Andrew Crocker,
        and Simão Melo de Sousa}

\IEEEtitleabstractindextext{%
\begin{abstract}

This paper presents Adamastor, a new low latency and scalable decentralized anonymous payment system, which is an extension of Ring Confidential Transactions (RingCT) that is compatible with consensus algorithms that use Delegated Proof of Stake (DPoS) as a defense mechanism against Sybil attacks. Adamastor also includes a new Decoy Selection Algorithm (DSA) that can be of independent interest, called SimpleDSA, a crucial aspect of protocols that use ring signatures to anonymize the sender. SimpleDSA offers security against homogeneity attacks and chain analysis. Moreover, it enables the pruning of spent outputs, addressing the issue of perpetual output growth commonly associated with such schemes. Adamastor is implemented and evaluated using the Narwhal consensus algorithm, demonstrating significantly lower latency compared to Proof of Work based cryptocurrencies. Adamastor also exhibits ample scalability, making it suitable for a decentralized and anonymous payment network.

\end{abstract}

}

\maketitle

\IEEEdisplaynontitleabstractindextext

% \fi
\IEEEpeerreviewmaketitle

\section{Introduction}\label{sec:introduction}

    Bitcoin \cite{bitcoin} has redefined the traditional financial systems by introducing an autonomous, secure, and borderless medium of exchange, with its innovative Nakamoto consensus \cite{nakamoto}. 
    Despite its time-tested reliability, it has proven to have some design limitations, it is computationally intensive and energy inefficient due to its use of Proof of Work (PoW) as a defense mechanism against sybil attacks, which can cause substantial delays in transaction confirmations and poses significant environmental concerns. Also, it is only pseudo-anonymous, since it is possible to trace an account of the ledger to a real world personal identity. Because of that, new protocols have been developed to solve these issues, however, to our knowledge none has been able to solve them all without sacrificing what motivated the research in this field in the first place: decentralization. 

This paper aims to explore and develop an innovative approach for addressing these critical challenges, striving for a balance between user anonymity, transaction efficiency and latency, system scalability, and decentralization. 

\subsection{Contributions}

In this work we have made the following significant contributions that can advance the state of the art in anonymous cryptocurrency protocols and blockchain technology:

\begin{itemize}
    \item Adamastor: We propose and developed a novel version of the Confidential Transaction protocol, which we have named Adamastor. This protocol maintains the privacy and anonymity features inherent to the original RingCT, while it has been specifically designed to be compatible with consensus algorithms that use DPoS. This work builds on but thoroughly extends our previous work on faster anonymous transaction for the RingCT protocol which can be found as a preprint    \cite{cryptoeprint:2020/1521}. In particular through extending the compatibility of RingCT to DPoS, we have fortified defenses against Sybil attacks, an enduring issue within decentralized systems. This modification allows us to broaden the applicable range of consensus mechanisms, moving beyond the traditional PoW, and offering potential enhancements in terms of latency and computational resource efficiency.

    \item SimpleDSA: Recognizing that decoy selection can be and has been a vector for deanonymization attacks in cryptocurrencies using RingCT, we have proposed an innovative and efficient DSA that can be of independent interest. This new algorithm effectively counters homogeneity attacks and chain analyses, while also offering a mechanism to identify spent outputs within the ledger and prune them accordingly. %This robust DSA fortifies the privacy-preserving characteristics of the proposed Adamastor, thus increasing its security profile.

    \item Implementation and evaluation with open-source code: To demonstrate the viability and effectiveness of our protocol, we have implemented Adamastor in Rust, a language renowned for its performance and safety, with Narwhal \cite{narwhal/tusk}, a state of the art consensus algorithm. Our comprehensive evaluations indicate that Adamastor significantly outperforms the existing state of the art anonymous decentralized payment systems in terms of speed and resource usage. In addition, we have made our implementation available as open-source, contributing a valuable tool for the broader research community and paving the way for future advancements.
\end{itemize}

%Through these contributions, we believe we have set a new standard for cryptocurrency protocols and decentralized systems, offering new avenues and opportunities for future research and applications.

\subsection{Related Work}

Initial attempts to provide anonymize Bitcoin involved mixing the unspent outputs to obfuscate the sender, like TumbleBit \cite{tumblebit} and Coinshuffle \cite{coinshuffle}, but they offer limited anonymity.

Because of that, new cryptocurrencies were developed, such as Monero. It was first described in \cite{cryptonote} and combines linkable ring signatures \cite{lrs} with confidential transactions \cite{coinjoin} into ring confidential transactions. Reference \cite{rct2} gives the first formal syntax of RingCT and improves it with a new version and \cite{rct3,omniring,arcturus,triptych} improve on the size and the efficiency of the linkable ring signature component and \cite{bulletproofs} on the range proof. Compatibility with smart-contracts was achieved in \cite{zether}.
    
    ZCash uses zero-knowledge Succinct Non-interactive AR-guments of Knowledge (zk-SNARKs) to construct a decentralised payment scheme \cite{zcash}, but requires a trusted setup. Since then, other zero-knowledge proofs for arithmetic circuits were developed by improving on efficiency, decreasing the amount of "trust" required \cite{sonic} or increasing the scope of use by adding smart contracts \cite{zexe}.

%groth, halo
    
Other relevant approaches to anonymous cryptocurrencies are Zerocoin \cite{zerocoin} its evolution Lelantus \cite{lelantus}, Dash, that uses masternodes to provide some degree of anonymity \cite{dash}, and Ouroboros Crypsinous, the first formal analysis of a privacy-preserving proof-of-stake blockchain protocol \cite{ouro_anon}. Mimblewimble \cite{mimblewimble} uses a different approach to achieve privacy and scalability. It leverages confidential transactions and combines it with a new technique that allows the pruning of old, unspent transaction outputs. 

Besides anonymous decentralized payment systems, there has also been research focusing on the development of frameworks to execute smart contracts privately, like Hawk \cite{hawk} and Zexe \cite{zexe}. Our work, Adamastor, addresses many of the limitations of these works, providing scalable, efficient, and privacy preserving transactions without relying on a trusted setup. It also introduces an effective Decoy Selection Algorithm to counter deanonymization attacks and prune spent outputs, offering an advancement in the secure implementation of privacy-preserving cryptocurrencies.

\subsection{Technical Overview}

In a general way, a transaction protocol defines a set of rules for transferring data (coins) between different parties in a network, which are recorded in a ledger. This typically follows the account model or the UTXO model \cite{utxo}, the one that we use in our protocol. These rules are enforced trough a consensus mechanism and are the following:
    
    \begin{itemize}
        \item Membership: Coins spent in a transaction must already exist in the ledger.
        
        \item Ownership: Coins must only be spent by its owner.
        
        \item Conservation of Value. The total amount of the coins consumed in a transaction must be equal to the amount of the coins created (plus an optional fee).

        \item No Double Spending: Coins spent in a transaction must not have been already spent.
    \end{itemize}
    
    A decentralized anonymous payment scheme tries to enforce these rules in a decentralized way and in a way that the amounts of the coins and the relationship between transactions are not revealed. 
    Our goal is to make it compatible with the DPoS family of consensus without breaking these rules, where the weight that a node has on the consensus of the network is proportional to its stake (total value of the coins) on the network and that stake can be delegated to another node, effectively transferring its weight on the consensus to that node.
     
    With this in mind, the concept of stake delegation is introduced by representing the amount of a coin as a Pedersen Commitment (PC), which is additive homomorphic. Both the receiver and the delegate of the coin need to know what is the amount of the coin, the first to be able to spend that coin and the second to be able to prove its delegated stake to the network.

    To do that, each output has two ciphertexts that encrypt the value of the corresponding coin, one for the receiver and the other for the delegate, where the key for decrypt the ciphertext is constructed with the transaction public key and the secret key of the receiver or the delegate. This can be seen as a three party Diffie-Hellman exchange \cite{diffie_hellman}, since there is
    in a three-way shared secret (the value of the coin) between the sender, the receiver and the delegate of the transaction output.

    All the inputs (outputs of a previous transaction) of a transaction need to have the same delegate and SimpleDSA chooses the decoys for a given spend transaction. A delegate is able to know the amount of an output that delegated to it, but it does not know what are the sender and receiver master addresses of a transaction.
    
    Since the PC is additive homomorphic and the consensus algorithm only needs to know the total amount of delegated stake to a given delegate, he can reveal the total value of the delegated coins without revealing the individual value of each coin in a reveal transaction.
    
    The owner of a coin can change its delegate anytime in a delegate transaction by producing a new PC together with a Non-Interactive Zero-Knowledge (NIZK) proof that the old and the new commitments have the same value, assuring that no new coins are created.

    This paper is organized as follows: Section II describes the building blocks that are used to develop the Adamastor protocol. Sections III defines it formally and section IV details SimpleDSA. In Section V, the implementation details and evaluation results of Adamastor are presented and the work is concluded in Section VI.
    
\section{Building Blocks}

    \subsection{Digital Signature Scheme}
    
    A Digital Signature scheme is composed of the following algorithms:

    \begin{itemize}

        \item {\cmss{Setup}}($1^{\lambda}$) $\rightarrow pp$. A probabilistic algorithm that receives as input a security parameter $\lambda$ and outputs public parameters $pp$.
        
        \item {\cmss{KeyGen}}($pp$) $\rightarrow (pk, sk)$. A probabilistic algorithm that receives as input $pp$, and outputs a public key $pk$ and a secret key $sk$.
    
        % format all algos
        \item {\cmss{Sign}}($sk, m$) $\rightarrow \sigma$. A probabilistic algorithm that receives as input a secret key $sk$ and a message $m$, and outputs a signature $\sigma$.
            
        \item {\cmss{Verify}}($pk, m, \sigma$) $\rightarrow 1/0$. A deterministic algorithm that receives as input a public key $pk$, a message $m$, and a signature $\sigma$, output “1” if the signature is valid and “0” if it is not.

    \end{itemize}

    \textbf{Existential Unforgeability under Chosen Message Attacks (EUF-CMA)}. A signature scheme is EUF-CMA secure if no efficient (polynomial-time) adversary can forge a valid signature for a new message, even if it has been able to see valid signatures for messages of its choosing.

    \subsection{Homomorphic Commitment Scheme}

    A Homomorphic Commitment scheme is composed of the following algorithms:

    \begin{itemize}
        \item {\cmss{Setup}}($1^{\lambda}$) $\rightarrow pp$. On input a security parameter $\lambda$, output public parameters $pp$, which include the generators $g$ and $h$. %We assume that $pp$ includes the descriptions of message space $M$, randomness space $R$, commitment space $C$ and the generators $g$ and $h$. pp will be used as implicit input of the following two algorithms.

        \item {\cmss{Com}}($m$) $\rightarrow c$. On input a message $m$, it chooses a random value $r$ and returns $c \leftarrow g^m * h^r$.

        \item {\cmss{Open}}($c, m, r$) $\rightarrow 1/0$. On input a commitment $c$, a message $m$ and a random value $r$, it computes $c' = g^m * h^r$ and returns 1 if $c = c'$ and 0 otherwise.
    \end{itemize}

    \textbf{Homomorphism.} For any commitment $c$, any $(m1, r1), (m2, r2) \in M x R$, we have $Com(m1, r1) + Com(m2, r2) = Com(m1 + m2, r1 + r2)$. 

    \subsection{Symmetric Encryption Scheme}

    A symmetric encryption scheme is a cryptographic system where the same key is used for both encryption and decryption of data. Formally, a symmetric encryption scheme consists of three main algorithms:

    \begin{itemize}
        \item {\cmss{Setup}}($1^{\lambda}$): on input a security parameter $\lambda$, output public parameters $pp$. % all algorithms have implicitly the public parameters pp as input
    
        \item {\cmss{KeyGen}}() $\rightarrow k$. This algorithm takes a security parameter as input and generates a secret key, denoted as $k$. The security parameter determines the key size and, consequently, the level of security provided by the encryption scheme.

        \item {\cmss{Enc}}($k, m$) $\rightarrow c$. The encryption algorithm takes the secret key $k$ and a plaintext message $m$ as inputs and outputs a ciphertext $c$. The ciphertext is the encrypted form of the plaintext message, and the process ensures that the information is secure from unauthorized access.

        \item {\cmss{Dec}}($k, c$) $\rightarrow m$. The decryption algorithm takes the secret key $k$ and a ciphertext $c$ as inputs and outputs the original plaintext message $m$. This process allows the intended receiver to recover the original information from the encrypted data.
    \end{itemize}

    \textbf{Indistinguishability under Chosen Plaintext Attack (IND-CPA).} A symmetric encryption scheme is IND-CPA secure if no polynomial-time adversary can distinguish between two ciphertexts that encrypt two plaintexts of its choosing.

    \subsection{Non-Interactive Zero-Knowledge Proof System}
        
        A NIZK in the random oracle model can be constructed by applying the Fiat-Shamir transform \cite{fs_transform} to the $\Sigma$-protocol by modeling a cryptographic hash function as random oracle and computing $e = \hs(m)$. This not only removes interaction, but also strengthens honest-verifier zero-knowledge to full zero-knowledge (against malicious verifiers). It is composed of the following three polynomial time algorithms.
        
        \begin{itemize}
            \item {\cmss{Setup}}($1^\lambda$) $\rightarrow pp$. A probabilistic algorithm that receives as input the security parameter $\lambda$ and outputs public parameters $pp$. 
                
            \item {\cmss{Prove}}($x, w$) $\rightarrow \pi$. A probabilistic algorithm that receives as input a statement-witness pair ($x, w$), and outputs a proof $\pi$.
            
            \item {\cmss{Verify}}($x, \pi$) $\rightarrow$ 1/0. A deterministic algorithm that receives as input a statement $x$ and a proof $\pi$, and outputs “0” if rejects and “1” if accepts.
        \end{itemize}

    \subsection{Decoy Selection Algorithm}

We adapt the definition of DSA to the context of an anonymous decentralized payment system, and it is composed of the following algorithms:

    \begin{itemize}
        \item {\cmss{Setup}}($1^{\lambda}$) $\rightarrow$ $pp$. It takes a security parameter $\lambda \in N$, and outputs the system parameters pp. All algorithms below have implicitly $pp$ as part of their inputs.

        \item {\cmss{selectDecoys}}($n, m, i$) $\rightarrow$ $pp$. It takes as input the order $i$ of the real output, the number of inputs in a transaction $m$ and the number of outputs in a transaction $n$.
    \end{itemize}

    \textbf{Security model.} A DSA is secure if it prevents the following attacks: 

    \begin{itemize}
    
        \item \textbf{Homogeneity Attack.} It is infeasible for an adversary A to determine the historical transaction $t'$ where the real input of a transaction $t$ originated ($t' < t)$, except with negligible probability in the security parameter $\lambda$.

        \item \textbf{Chain-reaction Analysis.} It is computationally infeasible for an adversary A to determine if an input (decoy) of a transaction $t$ has been previously consumed in a transaction $t'$ ($t' < t)$, except with negligible probability in the security parameter $\lambda$.

    \end{itemize}

    \subsection{Ring Confidential Transactions Protocol}

    A RingCT protocol \cite{rct3} consists of the following algorithms:

    \begin{itemize}
        \item {\cmss{Setup}}($1^{\lambda}$) $\rightarrow$ $pp$. It takes a security parameter $\lambda \in N$, and outputs the system parameters pp. All algorithms below implicitly contain these system parameters, $pp$, as part of their inputs.
        
        \item {\cmss{LongTermKeyGen}}() $\rightarrow (ltsk, ltpk)$. This algorithm outputs a long term secret key $ltsk$ and a long term public key $ltpk$.
        
        \item {\cmss{OneTimePKGen}}($ltpk$) $\rightarrow (pk, aux)$. Given a long term public key $ltpk$ as input it outputs a one-time public key $pk$ and the corresponding auxiliary information $aux$.
        
        \item {\cmss{OneTimeSKGen}}($pk, aux, ltsk$) $\rightarrow sl$. On input of a one-time public key $pk$, an auxiliary information $aux$ and a long term secret key $ltsk$, it outputs the one-time secret key $sk$. 

        \item {\cmss{Mint}}($pk, a$) $\rightarrow (cn, ck)$. Given a one-time public key $pk$ and an amount $a$ as input it outputs a coin $cn$ and the associated coin key $ck$.

        \item {\cmss{AccountGen}}($sk, pk, cn, ck, a$) $\rightarrow (act, ask)$. It takes as input a user key pair ($sk, pk$), a coin $cn$, a coin key $ck$ and an amount $a$. It returns $\perp$ if $ck$ is not the coin key of $cn$ with amount $a$. Otherwise, it outputs the account $act = (pk,cn)$ and the corresponding account secret key is $ask = (sk,ck,a)$.

        \item {\cmss{Spend}}($m, ask, A_{\text{in}}, O$) $\rightarrow (A_{\text{out}}, \pi_{\text{range}}, s, CK_{\text{out}})$. It takes as input a message $m$, an account secret key $ask$ and a set $A_{\text{in}}$, of input accounts in which the spending account is included and a set of output public keys, $O$, with the corresponding output amounts. It returns a set of output accounts $A_{\text{out}}$, a range proof $\pi$, a serial number $s$ and a set of output coin keys $CK_{\text{out}}$

        \item {\cmss{Verify}}($m, A_{\text{in}}, A_{\text{out}}, \pi, s$) $\rightarrow 1/0/ - 1$. It takes as input a message $m$, a set of input accounts $A_{\text{in}}$, a set of output accounts $A_{\text{out}}$, a range proof $\pi_{\text{range}}$ and a serial number $s$, the algorithm outputs -1 if the serial number was spent previously. Otherwise, it checks if the range proof $\pi_{\text{range}}$ is valid for the transaction, and outputs 1 or 0, meaning a valid or invalid spending respectively.

    \end{itemize}

    \textbf{Security model.} A RingCT protocol is secure if it is anonymous against receiver and ring insider, unforgeable, equivalent, linkable and non-slanderable \cite{rct3}.  

    \section{Definition of Adamastor}
    
    \subsection{Data Structures}

    \noindent \textbf{Ledger.} The Adamastor protocol operates on top of a ledger $L$, which is a publicly accessible and append-only database. At any given time $t$, all users have access to $L_t$, which is a sequence of transactions. If $t < t'$, state of $L_t$ is anterior to the state of $L_{t'}$.

    \noindent \textbf{Public parameters.} These include the group in which the algorithms perform operations, generators of the group, cryptographic hash functions and parameters regarding transactions, namely: 
    
    \begin{itemize}
        \item $a$, which specifies the maximum possible number of coins that the protocol can handle. Any balance and transfer must lie in the integer interval $a = [0, a\textsubscript{max}]$.
        
        \item $m$, the maximum number of input accounts used as decoys in a spend transaction.
        
        \item $n$, the maximum number of output coins of a spend transaction.
    \end{itemize}
    
    %\vspace{1mm}
    
    \noindent \textbf{Address.} A public key $pk$. It can be used to transfer the ownership of coins in a spend transaction or to delegate coins in a delegate transaction.
    
    %\vspace{1mm}
    
    \noindent \textbf{Coin.} A commitment of an amount $a$ and a randomness coin key $ck$.

    \noindent \textbf{Account.} A data structure composed of a coin $cn$, a one-time public key of the receiver $pk_{r}$, a public key of the delegate $pk_{d}$, a ciphertext for the receiver $c_r$, a ciphertext for the delegate $c_d$ and an auxiliary information $aux$. The auxiliary information $aux$ allow both the owner of the account (receiver) and its delegate to reveal the amount of the coin and also can be used by the receiver to compute the secret key $sk_r$ of the account, allowing it to spend it. The account secret key $ask$ is composed of the secret key $sk_r$, the amount $a$ and the coin key $ck$.
    
    \noindent \textbf{Delegate.} A delegate of a coin can stake its value and use it to participate in the consensus of the network to validate transactions. Its weight on the consensus is proportional to the total amount of delegated stake.
    
    %\vspace{1mm}
    
    \noindent \textbf{Transactions.} Structured blocks of data that are validated through the consensus mechanism and recorded on the ledger in the following four types of transactions.
    
    \begin{itemize}
        \item \noindent \textbf{Spend.} A spend transaction transfers the ownership of coins by creating new ones with the same value. All input and output coins of a spend transaction must have the same delegate so that the anonymity of the sender and the amounts is maintained.
        
        \item \noindent \textbf{Mint.} A mint transaction creates a new unspent coin through the consensus process.
        
        \item \noindent \textbf{Delegate.} A delegate transaction allows the owner of a coin to change its delegate. 
    
        \item \noindent \textbf{Reveal.} A reveal transaction is used by a receiver or a delegate to reveal the amount of an account. It can be used by a delegate to prove its total delegated stake to the network since the commitments used to represent coins are homomorphic.

    \end{itemize}
    
    \subsection{Algorithms}
    
    The Adamastor protocol is composed of the following polynomial-time algorithms.

   \begin{itemize}
    
        \item {\cmss{Setup}}($1^\lambda$) $\rightarrow pp$. On input a security parameter $\lambda$, it generates the public parameters $pp$ of Adamastor, which are implicit in the rest of the algorithms.
        
        \item {\cmss{LongTermKeyGen}}() $\rightarrow (ltpk, ltsk)$. Returns the long term public and secret keys $ltpk$ and $ltsk$. 

        \item {\cmss{OneTimePKGen}}($ltpk$) $\rightarrow (pk, aux)$. On input a long term public key $ltpk$, it outputs a one-time public key $pk$ and the auxiliary information $aux$.
        
        \item {\cmss{OneTimeSKGen}}($pk, aux, ltsk$) $\rightarrow sk$. On input a one-time public key $pk$, auxiliary information $aux$ and a long term secret key $ltsk$, it outputs the one-time secret key $sk$. 

        \item {\cmss{Mint}}($pk_r, a$) $\rightarrow (cn, ck)$. It takes as input the receiver public key $pk_r$ and an amount $a$, outputs a coin $cn$ and the associated coin key $ck$.

        \item {\cmss{AccountGen}}$(sk_r, pk_r, ltpk_d, cn, ck, a, aux) \rightarrow (act, ask)/ \perp$. It takes as input a receiver key pair ($sk_r, pk_r$), a delegate long term public key $ltpk_d$, a coin $cn$, a coin key $ck$, an amount $a$ and the auxiliary information $aux$. It returns $\perp$ if ck is not the coin key of $cn$ with amount $a$. Otherwise, it outputs the account $act = (pk_r, cn, pk_d, c_r, c_d, aux)$ and the corresponding account secret key is $ask = (sk_r, ck, a)$. %The auxiliary information $aux$ is used by the the receiver and the delegate to reveal the amount $a$ from the ciphertext $c_r$ and $c_d$, respectively. 
        
        \item {\cmss{Spend}}($m, i, ask, O, L$) $\rightarrow (A_{out}, \pi_{\text{range}}, s, CK_{out})/ \perp$. 

        It takes as input a message $m$, the index $i$ of the account to be spent in the ledger $L$, the corresponding account secret key $ask$, the set of output accounts with the corresponding output amounts $O$ and the ledger $L$. It outputs $\perp$ if the sum of output amounts in $O$ is different from the input amount or the $ask$ is not valid. Otherwise it returns the output accounts $A_{\text{out}}$, a range proof $\pi_{\text{range}}$, a serial number $s$ and a set of output coin keys $CK_{out}$.
        
        \item {\cmss{VerifySpend}}($m, A_{\text{in}}, A_{\text{out}}, \pi_{\text{range}}, s$) $\rightarrow$ 1/0/-1. it takes as input a message $m$, a set of input accounts $A_{\text{in}}$, a set of output accounts $A_{\text{out}}$, a range proof $\pi_{\text{range}}$ and a serial number $s$, the algorithm outputs -1 if the serial number has been spent previously. Otherwise, it checks if the range proof $\pi_{\text{range}}$ is valid for the transaction, and outputs 1 or 0, meaning a valid or invalid spending, respectively.
        
        \item {\cmss{Delegate}}($m, act, ask, ltpk_d'$) $\rightarrow (act', aux', \pi, \sigma)$. On input a message $m$, an account $act$ = ($pk_r, cn, ltpk_d, c_r, c_d, aux, ma$), the corresponding account secret key $ask$ = ($sk_r, ck, a$) and a new delegate long term public key $ltpk_d'$, it outputs a new account $act'$ = ($pk_r, cn', ltpk_d', c_r, c_d', aux'$), a proof of equivalence $\pi$ and a signature $\sigma$. %The auxiliary information $aux$ includes the transaction public key $txpk$ and the masked amount $ma$.
        
        \item {\cmss{VerifyDelegate}}($act, \sigma$) $\rightarrow$ 1/0. It receives an account $act$ and a signature $\sigma$ and returns "1" if it is is valid and "0" if not.
        
        \item {\cmss{Reveal}}($m, ltsk_{r/d}, act$) $\rightarrow (a, ck)$. On input a message $m$, a receiver or delegate long term secret key $ltsk_{r/d}$ and an account $act$, it returns the amount $a$ and the coin key $ck$ of the coin of that account. This algorithm serves as a proof of delegated stake for the delegate.
        
        \item {\cmss{VerifyReveal}}($cn, a, ck$) $\rightarrow$ 1/0. On input a transaction a coin $cn$, it outputs "1" if the amount $a$ and coin key $ck$ correspond to the coin $cn$ and "0" if not.
        
    \end{itemize}

    \subsection{Security Model}

    % Correctness

    We use the security model of \cite{rct3} and extend it to reflect the additions of Adamastor. The model consists of a polynomial time adversary that tries to break some security property by winning the corresponding security game. In the games, the adversary interacts with oracles that simulate the Adamastor protocol, which are the following:

    \begin{itemize}
        \item {\cmss{LongTermPKGen}}($i$). On input a query number $i$, it runs the algorithm {\cmss{LongTermKeyGen}}() $\rightarrow (ltsk_i, ltpk_i)$, adds ($i, ltsk_i, ltpk_i$) to an initially empty list and returns the long term public key $ltpk_i$.

        \item {\cmss{OneTimePkGen}}($i, j$). It retrieves ($i, ltsk_i, ltpk_i$) from $U'$ and runs {\cmss{OneTimePKGen}}($ltpk_i$) $\rightarrow (pk_{i,j}, R_{i,j})$. It adds ($i, j, *, pk_{i,j}, aux_{i,j}$) to an initially empty list and returns $pk_{i,j}$.

        \item {\cmss{OneTimeSkGen}}($i, j$). It retrieves $ltsk_i$ from $U$ and $pk_{i,j}$ and $aux_{i,j}$ from $U'$ and runs OneTimeSKGen($pk_{i,j}, aux_{i,j}, ltsk_i$) $\rightarrow sk_{i,j}$. It updates ($i, j, sk_{i,j}, pk_{i,j}, aux_{i,j}$) the corresponding list.

        \item {\cmss{ActGen}}($pk, a, ltpk_d$). On input a receiver public key $pk_r \in U$, an amount $a$ and a delegate long term public key $ltpk_d \in$ $U'$, it runs the algorithm {\cmss{Mint}}($a, pk_r, ltpk_d$) $\rightarrow (cn, ck)$, and outputs ($act, ck$) for address $pk_r$. It adds ($act, a, ck$) to an initially empty list $G$. If $(sk, pk) \in U$, then the associated secret key with account $act$ is $ask = (sk, ck, a$). It adds $act$ and ($act, ask$) to initially empty lists.

        \item {\cmss{Corrupt}}($act$): on input an account $act$, it retrieves $(act, ask) \in L$. It adds act to an initially empty list, and finally returns ask.

        \item {\cmss{Spend}}($m, i, O, L$): takes in a message $m$, the index $i$ of the account $act$ to be spent in the ledger $L$ and the output set $O$. It gets $act$ from the ledger, retrieves $(act, ask)$ from list I and runs {\cmss{Spend}}($m, i, ask, O, L$) $\rightarrow (A_{\text{out}}, \pi, s, CK_{\text{out}})$ and returns the result after adding it to an initially empty list.

        \item {\cmss{Delegate}}($act, ltpk_d'$). On input an account $act$ and a delegate long term public key $ltpk_d$, it retrieves the associated secret key $ask$ and runs the algorithm {\cmss{Delegate}}($act, ask, ltpk_d'$).

        \item {\cmss{Reveal}}($act$). On input an account $act$, it retrieves the associated receiver or delegate long term secret key and runs the algorithm {\cmss{ProveDelegatedStake}}($ltsk_{r/d}, act$).
    \end{itemize}

    %The definitions of the lists used by the oracles are the following:

    %\begin{itemize}
        %\item $U$: The list of user key pairs generated by the challenger.

        %\item $G$: The list of accounts, their balance and coin keys generated by Mint or Spend.

        %\item $I$: The list of accounts with public keys generated by the challenger.

        %\item $L$: The list of accounts and account secret keys with public keys generated by the challenger. 
        
        %\item $T$: The list of output generated by the Spend oracle with the corresponding input accounts. 
        
        %\item $C$: The list of corrupted  accounts, i.e., accounts that the adversary has access to the corresponding secret keys. 
    %\end{itemize}

    \newtheorem{definition}{Definition}
        
    \textbf{Anonymity against receiver}. The definition of anonymity against receiver ensures that, without the knowledge of the spender's account secret key or the delegate's secret key, the spender’s account is successfully hidden among all the honestly generated accounts, even if the output accounts and the output amounts are known. 
    
    \begin{definition}
        The Adamastor protocol is anonymous against receiver if a polynomial time adversary A has a negligible probability of winning the following security experiment with a challenger$\CH$:
    \end{definition}

    \begin{enumerate}
        \item \textbf{Setup:} $\CH$runs Adamastor.Setup($1^{\lambda}$) $\rightarrow pp$ and sends $pp$ to A.
    
        \item \textbf{Pre-challenge queries:} The adversary can query all previously defined oracles.

        \item \textbf{Challenge:} In the challenge phase, the adversary gives $(m,{pk_{k,i}}_{k \in [1,M], i \in [1,n]}, {pk_{out,j}}_{j \in [1,N]})$ to$\CH$. The adversary is given $A_{out}^*, \pi^* = (\pi_{range}, \sigma_{ring}), S^*$ and $A_{in}$, where $S^*$ is the set of serial numbers and $A_{in}$ is the set of input accounts.

        \item \textbf{Post-challenge queries:} After receiving the challenge, A can continue querying the same oracles, except corrupt accounts that were used in the challenge phase.

        \item \textbf{Guess:} The adversary outputs ($k^*, {ind_k}^*$). With probability 1/$n$, $k' = k^*$. The adversary wins with probability 1/$n$, if ${ind_k}^* = {i_k}^*$.
    \end{enumerate}

    \textbf{Anonymity against ring insider.} The definition of anonymity against ring insider ensures that, without the knowledge of the output accounts secret keys or the delegate's secret key, the spender's account is successfully hidden among all corrupted accounts.

    \begin{definition}
        The Adamastor protocol is anonymous against ring insider if a polynomial time adversary A has a negligible probability of winning the following security experiment with a challenger$\CH$:
    \end{definition}
    
    \begin{enumerate}
        \item \textbf{Setup}: $\CH$runs Adamastor.Setup($1^{\lambda}$) $\rightarrow pp$ and sends $pp$ to A.

        \item \textbf{Pre-challenge queries:} The adversary can query all previously defined oracles.

        \item \textbf{Challenge:} In the challenge phase, the adversary gives $(m,{act_{k,i}}_{k \in [1,M], i \in [1,n]}, {pk_{out,j}}_{j \in [1,N]})$ to$\CH$. The adversary is given $A_{out}^*, \pi^* = (\pi_{range}, \sigma_{ring}), S^*$ and $A_{in}$, where $S^*$ is the set of serial numbers and $A_{in}$ is the set of input accounts.

        \item Post-challenge queries: After receiving the challenge, A can continue querying the same oracles, including the  corrupt oracle, but the accounts that these oracles return cannot be corrupted, i.e., the inputs cannot be the accounts of the challenge.
        
        \item \textbf{Guess:} The adversary outputs ($k^*, {ind_k}^*$). The adversary wins with probability 1/($n-n^*$), if ${ind_k}^* = {i_k}^*$ and Ain $\cap$ C = $\emptyset$, where there are $n^*$ distinct values of $i$ such that $act_{k^*,i} \in C$.
    \end{enumerate}    

    \textbf{Unforgeability.} The unforgeability property ensures that no polynomial time adversary can forge a Spend transaction, a Delegate transaction or a Reveal transaction if all input accounts are not corrupted.
    %(corrupted).

    \begin{definition}
    The Adamastor protocol is unforgeable if a polynomial time adversary A has a negligible probability of winning the following security experiment with a challenger$\CH$:
    \end{definition}
    
    \begin{enumerate}
        \item \textbf{Setup}: $\CH$runs Adamastor.Setup($1^{\lambda}$) $\rightarrow pp$ and sends $pp$ to A.

        \item \textbf{Queries:} The adversary can query all previously defined oracles.

        \item \textbf{Forge:} A outputs a spend, delegate or reveal transaction such that no input account is corrupted and wins if the transaction is valid.
    \end{enumerate}    

        \textbf{Equivalence.} The equivalence property ensures that no polynomial time adversary can output a valid signature and coin keys where the input amount is different from the sum of the output amounts, even if the accounts are corrupted.

        \begin{definition}
            The Adamastor protocol is equivalent w.r.t. insider corruption if a polynomial time adversary A has a negligible probability of winning the following security experiment with a challenger$\CH$:
        \end{definition}
        
        \begin{enumerate}
            \item \textbf{Setup}: $\CH$runs Adamastor.Setup($1^{\lambda}$) $\rightarrow pp$ and sends $pp$ to A.
    
            \item \textbf{Queries:}: The adversary can query all previously defined oracles.
    
            \item \textbf{Output:} A outputs a spend transaction with an input amount different than the sum of the output amounts and wins if the transaction is valid.
        \end{enumerate} 

        \textbf{Linkability.} The linkability property ensures that no polynomial time adversary can output two valid spend transactions with the same serial number.

        \begin{definition}
            The Adamastor protocol is linkable w.r.t. insider corruption if a polynomial time adversary A has a negligible probability of winning the following security experiment with a challenger$\CH$:
        \end{definition}
        
        \begin{enumerate}
            \item \textbf{Setup}: $\CH$runs Adamastor.Setup($1^{\lambda}$) $\rightarrow pp$ and sends $pp$ to A.
    
            \item \textbf{Queries:} The adversary can query all previously defined oracles.
    
            \item \textbf{Output:} A outputs two spend transactions with the same serial number and wins if they are both valid.
        \end{enumerate} 
        
    \textbf{Non-Slanderability.} The definition of non-slanderabiliy ensures that no polynomial time adversary can make a spend transaction that is linkable with another spend transaction of an honest user, i.e., an adversary cannot produce a valid spend transaction with the same serial number of a previously honest spending. 

    \begin{definition}
            The Adamastor protocol is non-slandarable if a polynomial time adversary A has a negligible probability of winning the following security experiment with a challenger$\CH$:
        \end{definition}
        
        \begin{enumerate}
            \item \textbf{Setup}: $\CH$runs Adamastor.Setup($1^{\lambda}$) $\rightarrow pp$ and sends $pp$ to A.
    
            \item \textbf{Queries:} Same as in \textbf{Definition 1}.
    
            \item \textbf{Output:} A outputs a spend transaction with the same serial number of a previously honest spending and wins if the transaction is valid. 
        \end{enumerate} 

    \section{SimpleDSA: A Simple Decoy Selection Algorithm}

In the realm of cryptographic techniques for ensuring privacy and anonymity, one of the more interesting methods is the use of ring signatures. A ring signature allows for a user to sign a message on behalf of a group of users, such that an outsider cannot tell who among the group actually created the signature. The signer only needs the public keys of the other group members and doesn't need their cooperation. 

The effectiveness of ring signatures depends on the size of the ring and the proper selection of decoys, also known as 'ring members' or 'mixins'. These are the non-signing members whose public keys are included in the ring signature alongside the actual signer's key. The security and anonymity level that a ring signature can offer are largely influenced by how these decoys are chosen.

Choosing the right decoys in a ring signature is thus essential for achieving better anonymity. Ideally, we want a decoy selection algorithm that prevents the homogeneity attack and the chain-reaction analysis (as defined in the security model of Section 2), and that does not have a high computation complexity, since we want an user to be able to send and validate a transaction as fast as possible.

Monero started by having an algorithm that chose the decoys uniformly at random. However, that was found to be insecure \cite{monero2} and it was changed to follow a heuristic distribution based on past transactions, which is efficient and provides better anonymity, but does not guarantee the mentioned security properties. Recent work proposes a framework to find the decoys that guarantee those properties, but their most efficient instantiation has a computation complexity of $O(n^2)$ \cite{dsa}, since the selection depends on past transactions. 

Because of that, we propose a new decoy selection algorithm (Algorithm 1) that has only one predefined set of decoys for each real input in a transaction according to the number of inputs and outputs in the transaction and does not depend on past transactions. 

\begin{algorithm}
\DontPrintSemicolon
\SetAlgoLined
\KwResult{Select decoys}
\BlankLine

\SetKwFunction{selectdecoys}{select\_decoys}
\SetKwProg{myalg}{Algorithm}{}{}
\myalg{\selectdecoys{number\_of\_outputs, number\_of\_inputs, real\_input\_number}}{
    \KwIn{number\_of\_outputs, number\_of\_inputs, real\_input\_index}
    \KwOut{A shuffled subsequence}
    subsequences $\leftarrow$ []\;
    current\_subsequence $\leftarrow$ []\;
    v $\leftarrow$ 0\;
    \While{real\_input\_number not in current\_subsequence}{
        a $\leftarrow$ v\;
        \For{$i \leftarrow 1$ \KwTo number\_of\_inputs}{
            current\_subsequence.append(v)\;
            v $\leftarrow$ v + number\_of\_outputs\;
            \If{v $>$ real\_input\_index and len(current\_subsequence) == number\_of\_inputs}{
                a $\leftarrow$ a + 1\;
                v $\leftarrow$ a\;
                break\;
            }
        }
        \If{real\_input\_number in current\_subsequence}{
            break\;
        }
        subsequences.append(current\_subsequence)\;
        current\_subsequence $\leftarrow$ []\;
    }
    \Return current\_subsequence\;
}
\caption{SimpleDSA}
\end{algorithm}

For a better understanding we give an example with 3 as the number of inputs and 3 as the number of outputs with two possible scenarios:

\begin{itemize}
    \item tx3: $0,3,6 \rightarrow 10,11,12$

    \item tx4: $0,3,6 \rightarrow 13,14,15$

    \item tx5: $0,3,6 \rightarrow 16,17,18$
\end{itemize}

Scenario 1:

\begin{itemize}
    \item tx10: $10,11,12 \rightarrow 20,21,22$

    \item tx11: $10,11,12 \rightarrow 23,24,25$

    \item tx12: $10,11,12 \rightarrow 26,27,28$

    \item tx13: $10,11,14 \rightarrow 29,30,31$
\end{itemize}

Scenario 2:

\begin{itemize}
    \item tx10': $10,13,16 \rightarrow 20,21,22$

    \item tx11': $10,13,16 \rightarrow 23,24,25$

    \item tx12': $10,13,16 \rightarrow 26,27,28$

    \item tx13': $11,14,17 \rightarrow 29,30,31$
\end{itemize}

We can see that in scenario 1 we know that tx10, tx11, tx12 come from tx3 (homogeneity attack) and that output 14 is being spent in tx13 (chain-reaction analysis), while in scenario 2 (our algorithm) tx10',tx11' and tx12' can come from tx3, tx4 or tx5 and output 11, 14 or 17 are being spent in tx13'.

The time complexity of our algorithm is only $O(n)$, since the running time of the algorithm grows linearly with the size of the inputs.
The algorithm has the advantage that if all possible ring signatures of a batch have been constructed, we know that all the outputs in the batch have been spent. Because of that we can delete them from the ledger and prune it. The security proofs are presented in appendix 1.

The disadvantage of SimpleDSA is that the number of inputs and the number of outputs must be fixed, which should not be a problem in practice, since most transactions have only two outputs and the size of the ring should be irrelevant as long as it is big enough to provide a good level of anonymity.

\subsection{Security Proofs of SimpleDSA}
\label{AppendixAspdsa}

According to the security model of section 2, SimpleDSA is secure if it is (a) secure against homogeneity attacks and (b) secure against chain-reaction analysis.

\begin{theorem}
    SimpleDSA is secure against homogeneity attacks.
\end{theorem}

\begin{proof}
    Since each decoy in a valid selection comes from different past transaction, an adversary cannot deduce that a certain consumed coin comes from a specific past transaction.
\end{proof}

\begin{theorem}
    SimpleDSA is secure against Chain-reaction analysis.
\end{theorem}

\begin{proof}
    Chain-reaction analysis is when an adversary can eliminate decoys of a valid selection because it knows they were already spent. 
    However, in SimpleDSA this is not possible to do because each output has a predetermined selection of decoys, that is shared with the other decoys. This means that all rings will have the same elements and it is only possible to know if all outputs have been spent, but not each one individually.
\end{proof}

    \section{Instantiation of Adamastor}

    An instantiation of Adamastor from the building blocks presented in section 2 is now given, in the following way.
    
    \begin{itemize}
        \item Let {\cmss{HC}} = ({\cmss{Setup}}, {\cmss{Com}}, {\cmss{Open}}, {\cmss{Prove}}, {\cmss{Verify}}) be a Homomorphic Commitment scheme, instantiated by the Pedersen Commitment scheme \cite{pedersen}.

        \item Let {\cmss{SE}} = ({\cmss{Setup}}, {\cmss{KeyGen}}, {\cmss{Enc}}, {\cmss{Dec}}) be a Symmetric Encryption scheme, instantiated by the Advanced Encryption Standard scheme. 

        \item Let {\cmss{S}} = ({\cmss{Setup}}, {\cmss{KeyGen}}, {\cmss{Sign}}, {\cmss{Verify}}) be a Signature scheme, instantiated by the Schnorr Signature scheme \cite{schnorr_sig}.

        \item Let {\cmss{NIZK}} = ({\cmss{Setup}}, {\cmss{CRSGen}}, {\cmss{Prove}}, {\cmss{Verify}}) be a NIZK proof system for the equivalence of two discrete logarithms, formally defined as \{$(g_1,h_1,g_2,h_2) | \exists w \in Z_p \quad \text{s.t.} \quad \text{log}_{g1} h1 = w = \text{log}_{g2} h2$\}, instantiated by the Chaum-Pedersen scheme \cite{dlog_equiv}.

        % We describe our generic DCP construction using NIZK in the CRS model. The construction and security proof carries out naturally if using NIZK in the random oracle model instead.
    
        \item Let {\cmss{RCT}} = ({\cmss{Setup}}, {\cmss{KeyGen}}, {\cmss{Mint}}, {\cmss{Spend}}, {\cmss{AccountGen}}, {\cmss{Verify}}) be a Ring Confidential Transaction, instantiated by the RCT3.0 scheme \cite{rct3}.

        \item Let {\cmss{DSA}} = ({\cmss{Setup}}, {\cmss{KeyGen}}, {\cmss{Mint}}, {\cmss{Spend}}, {\cmss{AccountGen}}, {\cmss{Verify}}) be a Decoy Selection Algorithm, instantiated by the Simple Decoy Selection Algorithm.

        \item Let {\cmss{H}} be a cryptographic hash function {\cmss{H}} : $M$ x $G \rightarrow Z_p$, where $M$ denotes the message space, instantiated by SHA-256.
        
    \end{itemize}

    The protocol is composed of the following algorithms.

     \begin{itemize}
    
        \item {\cmss{Setup}}($1^\lambda$) $\rightarrow pp$. On input a security parameter $\lambda$, it generates the public parameters $pp$ for {\cmss{HC}}, {\cmss{SE}}, {\cmss{NIZK}}, {\cmss{S}}, {\cmss{RCT}} and {\cmss{DSA}}, and returns them. They are implicit in the rest of the algorithms.
        
        \item {\cmss{LongTermKeyGen}}() $\rightarrow (ltpk, ltsk)$. It runs {\cmss{RCT.LongTermKeyGen}}() $\rightarrow (ltpk, ltsk)$.

        \item {\cmss{OneTimePKGen}}($ltpk$) $\rightarrow (pk, aux)$. It runs {\cmss{RCT.OneTimePKGen}}($ltpk$) $\rightarrow (ltpk, ltsk)$.

        \item {\cmss{OneTimeSKGen}}($pk, aux, ltsk$) $\rightarrow (sk)$. It runs {\cmss{RCT.OneTimeSKGen}}($pk, aux, ltsk$) $\rightarrow (ltpk, ltsk)$.
        
        \item {\cmss{Mint}}($pk, a$) $\rightarrow (cn, ck)$. It takes as input a public key $pk$ and an amount $a$, and runs:

        \begin{itemize}
            \item Picks $ck \overset{R}{\leftarrow} Z_p$.

            \item {\cmss{HC.Com}}($a, ck$) $\rightarrow cn$.
        \end{itemize}
        
        It outputs a coin $cn$ for $pk$ as well as the associated coin key $ck$.

        \item {\cmss{AccountGen}}($sk_r, pk_r, ltpk_d, cn, ck, a, aux) \rightarrow (act, ask)/\perp$: it takes as input a user key pair ($sk, pk$), a coin $cn$, a coin key $ck$, an amount $a$, a delegate long-term public key $pk_d$. It runs:

        \begin{itemize}
            \item $pk_d \leftarrow ltpk_d.0$.

            %\item $aux_a \leftarrow R$.
        
            \item {\cmss{SE.Enc}}$(k_r, m) \rightarrow c_r$, where $k_r = sk_r * aux$.

            \item {\cmss{SE.Enc}}$(k_d, m) \rightarrow c_d$, where $k_d = sk_d * aux$.
        \end{itemize}

        It returns $\perp$ if $ck$ is not the coin key of $cn$ with amount $a$. Otherwise, it outputs the account $act = (pk, aux, cn, c_r, c_d)$ and the corresponding account secret key is $ask = (sk,ck,a)$.
    
        \item {\cmss{Spend}}($m, i, ask, O, L$) $\rightarrow (A_{\text{out}}, \pi_{\text{range}}, s, CK_{\text{out}})$.
        
        It takes as input some transaction string $m \in {0,1}^*$, the index $i$ of account $act$ in the ledger $L$ together with the corresponding account secret key $ask$, an arbitrary set $A_{\text{in}}$ of groups of input accounts containing $act$, a set $O$ of output public keys with the corresponding output amounts and a ledger with past confirmed transactions $L$. It runs:

        \begin{itemize}
            \item {\cmss{DSA.selectDecoys}}($n, m, i) \rightarrow js$, where $n$ and $m$ are the number of outputs and the number of inputs in the spend transaction, respectively, and they are both public parameters of Adamastor .

            \item {\cmss{L.get}}($i) \rightarrow act$.
        
            \item {\cmss{L.get}}($js) \rightarrow A_{\text{in}}$.
        
            \item {\cmss{RCT.Spend}}$(m, ask, A_{\text{in}}, O) \rightarrow (A_{\text{out}}, \pi_{\text{range}}, s, CK_{\text{out}})$.
        \end{itemize}

        It outputs a set of output accounts $A_{\text{out}}$, a range proof $\pi$, a serial number $s$ and a set of output coin keys $CK_{\text{out}}$.
    
        \item {\cmss{VerifySpend}}$(m, A_{\text{in}}, A_{\text{out}}, \pi_{\text{range}}, s) \rightarrow 1/0/ - 1$: it takes as input a message $m$, a set of input accounts $A_{\text{in}}$, a set of output accounts $A_{\text{out}}$, a range proof $\pi_{\text{range}}$ and a serial number $s$.
        
        It runs {\cmss{RCT.Verify}}($m, A_{in}, A_{out}, \pi_{\text{range}}, s$) and outputs -1 if the serial number was already spent previously. Otherwise, it checks if the range proof $\pi_{\text{range}}$ is valid for the transaction, and outputs 1 or 0, meaning a valid or invalid spending respectively. 
        
        \item {\cmss{Delegate}}$(m, act, ask, ltpk_d') \rightarrow (\pi_{\text{equal}}, \sigma)$. On input a message $m$, an account $act$, the secret key of the account $ask$ and the new delegate public key $ltpk'd$, it runs:

        \begin{itemize}
            \item $pk_d' \leftarrow ltpk_d'.0$.
        
            \item Picks a random new coin key $ck'$ and computes the new coin $cn' \leftarrow$ {\cmss{HC.Com.}}($act.a, ck'$).
        
            \item {\cmss{NIZK.Prove}}$((h, cn, h, cn'), (ck - ck')) \rightarrow \pi_{\text{equal}}$.

            \item {\cmss{SE.Enc}}$(k_r, act.c_r) \rightarrow m$, where $k_r = act.pk_r * act.aux$.

            \item {\cmss{SE.Enc}}$(k_d, m) \rightarrow c_d$, where $k_d = pk_d' * act.aux$.

            \item {\cmss{S.Sign}}$(sk, m) \rightarrow \sigma$.
        \end{itemize}

        It returns the equivalence proof $\pi_{\text{equal}}$, the new delegate ciphertext $c_d'$ and the signature $\sigma$.

        \item {\cmss{VerifyDelegate}}$(act, act', \pi_{\text{equal}}, \sigma) \rightarrow 1/0$. On input two accounts $act$ and $act'$, an equivalence proof $\pi_{\text{equal}}$ and a signature $\sigma$ it runs:

        \begin{itemize}
            \item {\cmss{NIZK.Verify}}$((h, act.cn, h, act'.cn), \pi_{\text{equal}}) \rightarrow 1/0$.

            \item {\cmss{S.Verify}}$(act.pk, \pi_{\text{equal}}, \sigma) \rightarrow 1/0$.
        \end{itemize}
        
        It returns "1" if both return "1" and "0" otherwise.
        
        \item {\cmss{Reveal}}$(m, act, sk_{r/d}, aux) \rightarrow (a, ck)$. On input an account $act$, a receiver or delegate one-time secret key $sk_{r/d}$, it runs: 

        \begin{itemize}

            \item {\cmss{SE.Dec}}($k_{r/d}, act.c_{r/d}$) $\rightarrow m$, where $k_{r/d} = aux * sk_{r/d}$.

            \item {\cmss{H}}("key", $m$) $\rightarrow ck$.

            \item $ma$ $\oplus_8$ {\cmss{H}}("amount", $m$) $\rightarrow a$. The masked amount $ma$ is also included in the transaction by the sender.

            \item {\cmss{S.Sign}}$(sk_{r/d}, m) \rightarrow \sigma$.

            % add hash function to preliminares 
        \end{itemize}

        It returns the amount $a$ and the coin key $ck$ together with the signature $\sigma$.
        
        \item {\cmss{VerifyReveal}}($cn, a, ck, pk_d, \sigma) \rightarrow 1/0$. On input a coin $cn$, an amount $a$, a coin key $ck$, a delegate public key $pk_d$ and a signature $\sigma$, it runs:

        \begin{itemize}
            \item {\cmss{HC.Verify}}($act.cn, a, ck) \rightarrow 1/0$.

            \item {\cmss{S.Verify}}$(pk_d, m, \sigma) \rightarrow 1/0$.
        \end{itemize}
        
        It returns 1 if both algorithms return 1, and 0 otherwise.
    \end{itemize}

    \subsection{Security Proofs}

    We begin by explaining the proofs informally  and defer the formal proofs to the next section.

    The intuition for the anonymity, linkability, equivalence and non-slanderability is the following:
    
    Since the underlying RCT component defines these properties in the same way, i.e., the output of the adversary is the same,  and these are proven secure, we want to reduce the security experiment of Adamastor to the one of RingCT. If we can make this reduction so that the view of the adversary A is not altered, the probability that A has of winning will be the same in both experiments. Since this probability is negligible for the experiment of the RingCT's property, it will also be negligible for the experiment of the Adamastor's property, proving its safety.
    
    Since the only difference between the two security experiments is the access to new oracles that mimic the added functionality to the protocol, we can simulate these oracles so that they produce the same results but they do not use the real added underlying algorithms because the challenger $\CH$ of the experiment of RingCT does not have access to those algorithms. 

    The property of unforgeability is different because it captures the unforgeability of all transactions (Spend, Delegate and Reveal) and some of the output that A can produce is directly related to the added functionality of Adamastor. 
    
    If the output is a Spend transaction, we can follow the same idea as the other security properties. If the output is a Delegate or Reveal transaction, we cannot do that because RingCT does not have that functionality, so we reduce the security experiment to some underlying secure problem where the output of the Adamastor experiment can be used to try to break its security. 
    
    In the case of the Delegate transaction it is the Schnorr Signature scheme and in the case of the Reveal transaction it is the hiding property of the Pedersen Commitment. To do that we need to be able to simulate all the other components of Adamastor.

\subsection{Security Proofs of Adamastor}

\begin{theorem}
    Assuming the security of the \textit{S} component, the security of the {\cmss{RCT}} component, the security of the {\cmss{DSA}} component, the security of the {\cmss{SE}} component, the security of the {\cmss{HC}} component and the zero-knowledge property of {\cmss{NIZK}}, our {\cmss{Adamastor}} construction is anonymous against ring insider.
\end{theorem}

\begin{proof}
    We proceed via a sequence of games. Let $S_i$ be the probability that $A$ wins in Game $i$.

    \textbf{Game 0.} The real experiment for anonymity against ring insider as described in the security model of section 3.3.

    \textbf{Game 1.} The same as Game 0 except that the oracles of {\cmss{Adamastor}} that are not present in the security model of {\cmss{RCT}} are simulated by $\CH$ instead of using the real algorithms, in the following way:

    \begin{itemize}
        %\item ActGen: $\CH$picks a random amount $a'$ and uses it instead of the real amount $a$ given by the A.
    
        \item Delegate: $\CH$ can easily simulate the {\cmss{NIZK}} and the encryption part. 
        The signing part is simulated as described in the security proof of the Integrated Signature Encryption scheme in \cite{pgc}.

        \item Reveal: $\CH$ stores retrieves the amount $a$ and coin key $ck$ previously stored when running the {\cmss{ActGen}} oracle.
    \end{itemize}

    $\A$'s view of Game 1 is identical to Game 0 because $\CH$ perfectly mimics the oracles that use the simulated components of {\cmss{Adamastor}} and those components are secure. So, we have:

    $$| Pr[G_1] - Pr[G_0]| \leq negl(\lambda)$$

    \begin{lemma}
        Assuming the anonymity against ring insider of the {\cmss{RCT}} component, Pr[$G_1$] is negligible in $\lambda$ for any {\cmss{polynomial time}} adversary A.
    \end{lemma} 
    
    \begin{proof}
        We prove this lemma by showing that if there exists a {\cmss{polynomial time}} adversary $\A$ that has non-negligible advantage in Game 1, we can build a {\cmss{polynomial time}} adversary $\B$ that breaks the anonymity against ring insider of the {\cmss{RCT}} component with the same advantage since $\B$ can perfectly simulate Game 1 by forwarding the queries related to {\cmss{RCT}} to its own challenger and simulating the rest of the components of {\cmss{Adamastor}}. The guess of $\A$ is used by $\B$ in its own game and wins with the same probability as $\A$. 
    
        The security of the {\cmss{DSA}} component ensures an adversary cannot increase its probability of winning by eliminating decoys from the ring signature of a spend transaction.
    \end{proof} 

    The proof of the theorem follows directly from the lemma.
\end{proof}

\begin{theorem}
    Assuming the security of the {\cmss{S}} component, the security of the {\cmss{RCT}} component, the security of the {\cmss{SE}} component, the security of the {\cmss{HC}} component and the zero-knowledge property of {\cmss{NIZK}}, our {\cmss{Adamastor}} construction is anonymous against receiver.
\end{theorem}

\begin{proof}
    We proceed via a sequence of games. Let $S_i$ be the probability that $A$ wins in Game $i$.

    \textbf{Game 0.} The real experiment for anonymity against receiver as described in the security model of section 4.3.

    \textbf{Game 1.} The same as Game 0 except that the oracles of {\cmss{Adamastor}} that are not present in the security model of {\cmss{RCT}} are simulated by $\CH$ instead of using the real algorithms, in the following way:

    \begin{itemize}
    
        \item Delegate: $\CH$ can easily simulate the {\cmss{NIZK}} and the encryption part. The signing part is simulated as described in the security proof of the Integrated Signature Encryption scheme in \cite{pgc}.

        \item Reveal: $\CH$ stores retrieves the amount $a$ and coin key $ck$ previously stored when running the {\cmss{ActGen}} oracle.
    \end{itemize}

    $\A$'s view of Game 1 is identical to Game 0 because $\CH$ perfectly mimics the oracles that use the simulated components of {\cmss{Adamastor}} and those components are secure. So, we have:

    $$| Pr[G_1] - Pr[G_0]| \leq negl(\lambda)$$

    \begin{lemma}
        Assuming the anonymity against ring insider of the {\cmss{RCT}} component, Pr[$G_1$] is negligible in $\lambda$ for any {\cmss{polynomial time}} adversary A.
    \end{lemma} 
    
    \begin{proof}
        We prove this lemma by showing that if there exists a {\cmss{polynomial time}} adversary $\A$ has non-negligible advantage in Game 1, we can build a {\cmss{polynomial time}} adversary $\B$ that breaks the anonymity against ring insider of the {\cmss{RCT}} component with the same advantage since $\B$ can perfectly simulate Game 1 by forwarding the queries related to {\cmss{RCT}} to its own challenger and simulating the rest of the components of {\cmss{Adamastor}}. The guess of $\A$ is used by $\B$ in its own game and wins with the same probability as $\A$.

        The security of the {\cmss{DSA}} component ensures an adversary cannot increase its probability of winning by eliminating decoys from the ring signature of a spend transaction.
    \end{proof} 

    The proof of the theorem follows directly from the lemma.
\end{proof}

\begin{theorem}
    Assuming the security of the S component, the security of the RCT component, the security of the SE component, the security of the HC component and the zero-knowledge property of NIZK, our Adamastor construction satisfies unforgeability.
\end{theorem}

\begin{proof}
    We proceed via a sequence of games. Let Si be the probability that A wins in Game i. 
    
    \textbf{Game 0.} The real experiment for unforgeability as described in the security model of section 3.3.

    \textbf{Game 1.} The same as game 0 except for:

    \begin{enumerate}
        \setcounter{enumi}{1}
        
        \item Queries: in one query to the {\cmss{ActGen}} oracle $\CH$ picks a random public key $pk^*$ without knowing the corresponding secret key $sk^*$ and the coin $cn^*$ is a random commitment. It outputs the account $act^*$. 
        
        If the Spend oracle is queried with an account $act$ whose account secret key $ack$ is not known to$\CH$, the Spend transaction is simulated as in the proof of unforgeability of \cite{rct3} (the ciphertexts are picked at random) and the Delegate and Reveal oracles are simulated as in the previous anonymity proof.
    \end{enumerate}
    
    A’s view in Game 0 and Game 1 are identical because $\CH$can perfectly mimic the Spend, Delegate and Reveal oracles.

    $$| Pr[G_1] - Pr[G_0]| \leq negl(\lambda)$$

    \begin{lemma}
        Assuming the EUF-CMA security of the Schnorr Signature scheme, the hiding property of the Pedersen Commitment scheme and the unforgeability of the RCT3.0 scheme, Pr[$G_1$] is negligible in $\lambda$ for any polynomial time adversary $\A$.
    \end{lemma} 
    
    \begin{proof}
        We prove this claim by showing that if there exists a polynomial time adversary A has non-negligible advantage in Game 1, we can build a polynomial time adversary B that breaks either the EUF-CMA security of the Schnorr Signature, the hiding property of the Pedersen Commitment scheme or the unforgeability of the RCT3.0 scheme with the same advantage if the forge of A is a Delegate, a Reveal or a Spend, respectively. 
    \end{proof} 

    The proof of the theorem follows directly from the lemma.
\end{proof}

\begin{theorem}
    Assuming the security of the S component, the security of the RCT component, the security of the SE component, the security of the HC component and the zero-knowledge property of NIZK, our Adamastor construction satisfies equivalence w.r.t. ring insider.
\end{theorem}

\begin{proof}
    We proceed via a sequence of games. Let $S_i$ be the probability that $A$ wins in Game $i$.

    \textbf{Game 0.} The real experiment for anonymity against ring insider as described in the security model of section 3.3.

    \textbf{Game 1.} The same as Game 0 except that the oracles of {\cmss{Adamastor}} that are not present in the security model of {\cmss{RCT}} are simulated by $\CH$ instead of using the real algorithms, in the following way:

    \begin{itemize}
    
        \item Delegate: $\CH$ can easily simulate the {\cmss{NIZK}} and the encryption part. The signing part is simulated as described in the security proof of the Integrated Signature Encryption scheme in \cite{pgc}.

        \item Reveal: $\CH$ stores retrieves the amount $a$ and coin key $ck$ previously stored when running the {\cmss{ActGen}} oracle.
    \end{itemize}

    $\A$'s view of Game 1 is identical to Game 0 because $\CH$ perfectly mimics the oracles that use the simulated components of {\cmss{Adamastor}} and those components are secure. So, we have:

    $$| Pr[G_1] - Pr[G_0]| \leq negl(\lambda)$$

    \begin{lemma} 
    Assuming the anonymity against ring insider of the {\cmss{RCT}} component, Pr[$G_1$] is negligible in $\lambda$ for any {\cmss{polynomial time}} adversary A.
    \end{lemma}
    
    \begin{proof} We prove this lemma by showing that if there exists a {\cmss{polynomial time}} adversary $\A$ has non-negligible advantage in Game 1, we can build a {\cmss{polynomial time}} adversary $\B$ that breaks the anonymity against ring insider of the {\cmss{RCT}} component with the same advantage since $\B$ can perfectly simulate Game 1 by forwarding the queries related to {\cmss{RCT}} to its own challenger and simulating the rest of the components of {\cmss{Adamastor}}. The guess of $\A$ is used by $\B$ in its own game and wins with the same probability as $\A$. 
    \end{proof}

    The proof of the theorem follows directly from the lemma.

    \end{proof}

\begin{theorem}
    Assuming the security of the S component, the security of the RCT component, the security of the SE component, the security of the HC component and the zero-knowledge property of NIZK, our Adamastor construction satisfies linkability w.r.t. ring insider.
\end{theorem}

\begin{proof}
    We proceed via a sequence of games. Let $S_i$ be the probability that $A$ wins in Game $i$.

    \textbf{Game 0.} The real experiment for anonymity against ring insider as described in the security model of section 3.3.

    \textbf{Game 1.} The same as Game 0 except that the oracles of {\cmss{Adamastor}} that are not present in the security model of {\cmss{RCT}} are simulated by $\CH$ instead of using the real algorithms, in the following way:

    \begin{itemize}
    
        \item Delegate: $\CH$ can easily simulate the {\cmss{NIZK}} and the encryption part. The signing part is simulated as described in the security proof of the Integrated Signature Encryption scheme in \cite{pgc}.

        \item Reveal: $\CH$ stores retrieves the amount $a$ and coin key $ck$ previously stored when running the {\cmss{ActGen}} oracle.
    \end{itemize}

    $\A$'s view of Game 1 is identical to Game 0 because $\CH$ perfectly mimics the oracles that use the simulated components of {\cmss{Adamastor}} and those components are secure. So, we have:

    $$| Pr[G_1] - Pr[G_0]| \leq negl(\lambda)$$

    \begin{lemma}
        Assuming the anonymity against ring insider of the {\cmss{RCT}} component, Pr[$G_1$] is negligible in $\lambda$ for any {\cmss{polynomial time}} adversary A.
    \end{lemma} 
    
    \begin{proof}
        We prove this lemma by showing that if there exists a {\cmss{polynomial time}} adversary $\A$ has non-negligible advantage in Game 1, we can build a {\cmss{polynomial time}} adversary $\B$ that breaks the anonymity against ring insider of the {\cmss{RCT}} component with the same advantage since $\B$ can perfectly simulate Game 1 by forwarding the queries related to {\cmss{RCT}} to its own challenger and simulating the rest of the components of {\cmss{Adamastor}}. The guess of $\A$ is used by $\B$ in its own game and wins with the same probability as $\A$.
    \end{proof} 

    The proof of the theorem follows directly from the lemma.
\end{proof}

\begin{theorem}
    Assuming the security of the S component, the security of the RCT component, the security of the SE component, the security of the HC component and the zero-knowledge property of NIZK, our Adamastor construction satisfies non-slanderability.
\end{theorem}

\begin{proof}
    We proceed via a sequence of games. Let $S_i$ be the probability that $A$ wins in Game $i$.

    \textbf{Game 0.} The real experiment for anonymity against ring insider as described in the security model of section 3.3.

    \textbf{Game 1.} The same as Game 0 except that the oracles of {\cmss{Adamastor}} that are not present in the security model of {\cmss{RCT}} are simulated by $\CH$ instead of using the real algorithms, in the following way:

    \begin{itemize}
    
        \item Delegate: $\CH$ can easily simulate the {\cmss{NIZK}} and the encryption part. The signing part is simulated as described in the security proof of the Integrated Signature Encryption scheme in \cite{pgc}.

        \item Reveal: $\CH$ stores retrieves the amount $a$ and coin key $ck$ previously stored when running the {\cmss{ActGen}} oracle.
    \end{itemize}

    $\A$'s view of Game 1 is identical to Game 0 because $\CH$ perfectly mimics the oracles that use the simulated components of {\cmss{Adamastor}} and those components are secure. So, we have:

    $$| Pr[G_1] - Pr[G_0]| \leq negl(\lambda)$$

    \begin{lemma}
        Assuming the anonymity against ring insider of the {\cmss{RCT}} component, Pr[$G_1$] is negligible in $\lambda$ for any {\cmss{polynomial time}} adversary A.
    \end{lemma} 
    
    \begin{proof}
        We prove this lemma by showing that if there exists a {\cmss{polynomial time}} adversary $\A$ has non-negligible advantage in Game 1, we can build a {\cmss{polynomial time}} adversary $\B$ that breaks the anonymity against ring insider of the {\cmss{RCT}} component with the same advantage since $\B$ can perfectly simulate Game 1 by forwarding the queries related to {\cmss{RCT}} to its own challenger and simulating the rest of the components of {\cmss{Adamastor}}. The guess of $\A$ is used by $\B$ in its own game and wins with the same probability as $\A$.
    \end{proof}

    The proof of the theorem follows directly from the lemma.
\end{proof}

    \subsection{Implementation and Evaluation}

    \begin{figure}[htbp]
    \centering
    \includegraphics[width=\linewidth]{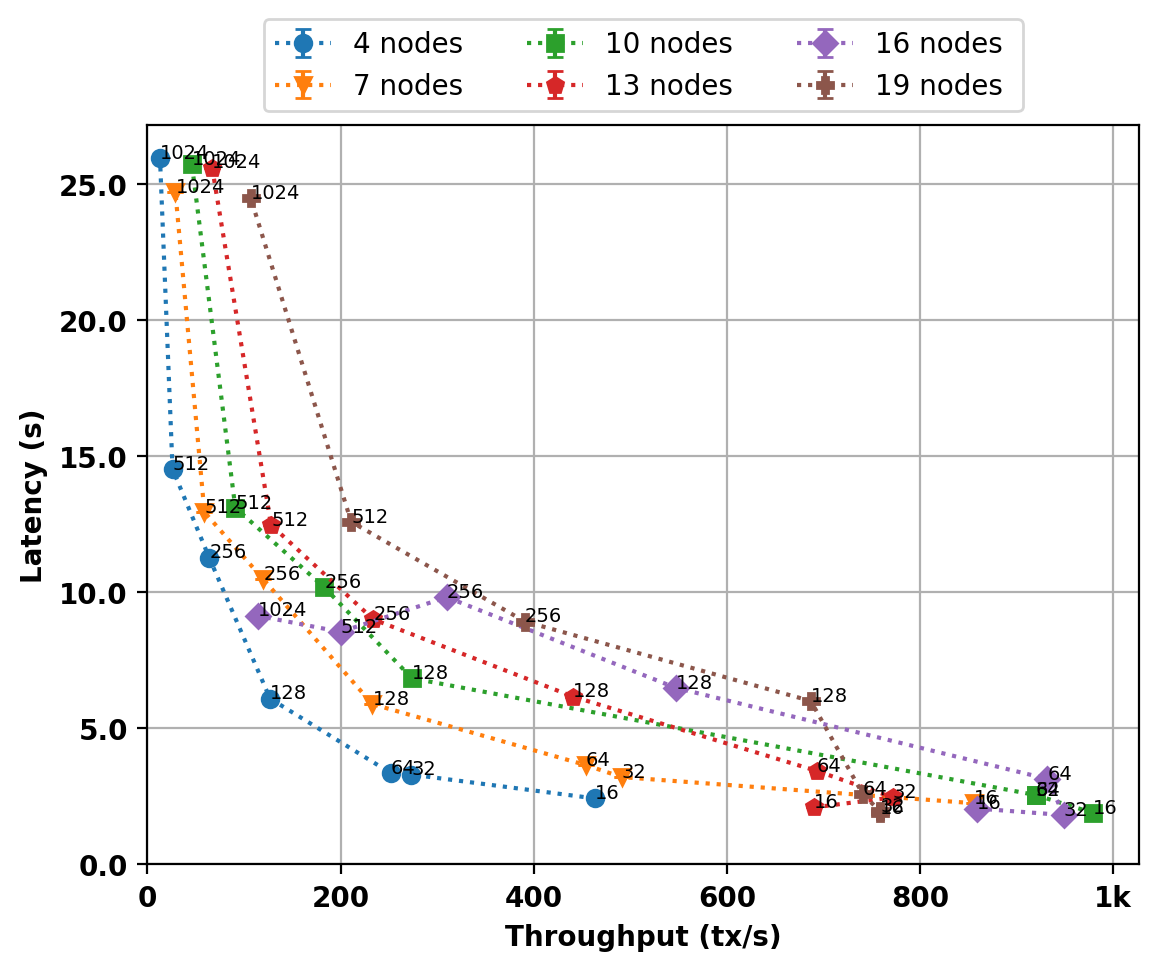}
    \caption{Benchmarking results of Adamastor}
    \label{fig:enter-label}
    \end{figure}

    We implemented a prototype of Adamastor in Rust \footnote{\url{https://github.com/Fiono11/Adamastor}} with the consensus algorithm of \cite{narwhal/tusk}. It was bench-marked using the Microsoft Azure platform \footnote{\url{https://azure.microsoft.com/en-us/free/students/}}. 
    Each node corresponds is a Standard\_DS1\_v2 instance, the instance with the lowest specs available with an Intel(R) Xeon(R) CPU E5-2673 v3 @ 2.40GHz processor and 3.5 GiB of memory.

    The results are plotted in Figure 1.  
    The numbers next to each data point represent the number of members used in the ring signature of a transaction (16, 32, 64, 128, 256, 512 and 1024). The greater the ring size the stronger the level of anonymity, but also the higher the latency and the lower the throughput. 
    
    Interestingly, we can observe that the latency does not change much with the number of nodes, and the throughput even increases with the increase of the number of nodes. The first is explained by the parallelism in the transaction verification between nodes and the second by how the broadcasting is done in the test, where in each run 1000 transactions are broadcast per second divided by the number of nodes and the consensus algorithm used, which batches votes. Each transaction has the number of members of the ring signature as inputs and only one output, since our protocol does not need fees.

    % 1 input, 1 output txs 
    
    \begin{table}[h]
    \centering
    \begin{tabular}{|l|c|c|}
    \hline
    & \textbf{Size ($\mathbb{F}$)} & \textbf{Size ($\mathbb{G}$)} \\
    \hline
    
    \textbf{Spend} & & $\Delta 3n + \Delta 3t$ \\
    \hline
    \textbf{Mint} & & $\Delta 3$ \\
    \hline
    \textbf{Delegate} & 4 & 3 \\
    \hline
    \textbf{Reveal} & 4 & 2 \\ % 1 NIZKdec for each ciphertext
    \hline
    \end{tabular}
    
    %\vspace{1mm}

    \caption{Proof sizes of Adamastor transactions.}
    \label{table:size}
    \end{table}
    
    A spend transaction of Adamastor has an additional overhead of 3$n + 3t$ and 3 elements of $\mathbb{G}$ when compared to the underlying RingCT scheme, respectively, corresponding to the ciphertext of the PKE scheme and the NIZK proof of encryption.

    Delegation and PoDS transactions have proof size of $4 \mathbb{F} + 3 \mathbb{G}$ and $4 \mathbb{F} + 2 \mathbb{G}$, respectively.

\section{Conclusion}

In conclusion, this work introduces Adamastor, an innovative decentralized anonymous payment system. By extending the capabilities of RingCT and incorporating a novel DSA, called SimpleDSA, we have developed a system that not only ensures robust security against homogeneity attacks and chain analysis, but also mitigates the problem of ever-increasing outputs in ring signature-based protocols. Our evaluation of Adamastor reveals remarkable improvements in latency while maintaining scalability and anonymity, setting a new standard in the realm of decentralized  that use DPoS and paving the way for further advances in this fast-evolving field.

In Adamastor, delegating accounts to other nodes decreases the level of anonymity of the sender and the receiver, since the delegate knows the amount transferred in a spend transaction. However, the delegate still does not know the long term public key associated of both the sending and the receiving account. Moreover, you can always be your own delegate and maintain the level of anonymity of RingCT. Because of this, we think that the advantages of Adamastor greatly outweigh its disadvantages, namely the increase in the complexity of the protocol and the implementation.

%\ifCLASSOPTIONcaptionsoff
  %\newpage
%\fi



%\bibliographystyle{IEEEtran}
%\begin{thebibliography}{1}

  %0.5em minus 0.4em\relax Harlow, England: Addison-Wesley, 1999.

%\end{thebibliography}

\bibliographystyle{IEEEtran}
\bibliography{references}


%\end{IEEEbiography}


%\newpage

%\vfill


%\appendices
%\include{IEEEtran/appendices}


    \end{document}